\documentclass{article}

\PassOptionsToPackage{numbers}{natbib}

\usepackage{graphicx}
\usepackage{subcaption} 

\usepackage[preprint]{neurips_2024}

\usepackage[utf8]{inputenc} 
\usepackage[T1]{fontenc}    
\usepackage{hyperref}       
\usepackage{url}            
\usepackage{booktabs}       
\usepackage{amsfonts}       
\usepackage{nicefrac}       
\usepackage{microtype}      
\usepackage{amsmath}
\usepackage{textcomp}

\usepackage{geometry}
\usepackage{lipsum,tabularx}
\hypersetup{
    colorlinks=true,
    linkcolor=blue,
    filecolor=magenta,      
    urlcolor=cyan,
    pdftitle={Overleaf Example},
    pdfpagemode=FullScreen,
    }

\title{Development of Machine Learning Classifiers for Blood-based
  Diagnosis and Prognosis of Suspected Acute Infections and Sepsis}

\author{%
  Ljubomir Buturovi\'c\thanks{lbuturovic@inflammatix.com. Authors are at Inflammatix Inc., Sunnyvale, CA, United States.}
  \And Michael Mayhew \And Roland Luethy \And
  Kirindi Choi \And Uro\v{s} Midi\'c \And Nandita Damaraju \And Yehudit
  Hasin-Brumshtein \And Amitesh Pratap \And Rhys M. Adams \And Joao
  Fonseca \And Ambika Srinath \And Paul Fleming \And Claudia Pereira
  \And Oliver Liesenfeld \And Purvesh Khatri \And Timothy Sweeney
}

\begin{document}

\maketitle

\begin{abstract}

  We applied machine learning to the unmet medical need of rapid and
  accurate diagnosis and prognosis of acute infections and sepsis in
  emergency departments. Our solution consists of a
  Myrna\textsuperscript{\texttrademark} Instrument and embedded
  TriVerity\textsuperscript{\texttrademark} classifiers. The
  instrument measures abundances of 29 messenger RNAs in patient's
  blood, subsequently used as features for machine learning. The
  classifiers convert the input features to an intuitive test report
  comprising the separate likelihoods of (1) a bacterial infection (2)
  a viral infection, and (3) severity (need for Intensive Care
  Unit-level care). In internal validation, the system achieved AUROC
  = 0.83 on the three-class disease diagnosis (bacterial, viral, or
  non-infected) and AUROC = 0.77 on binary prognosis of disease
  severity. The Myrna, TriVerity system was granted breakthrough
  device designation by the United States Food and Drug Administration
  (FDA). This engineering manuscript teaches the standard and novel
  machine learning methods used to translate an academic research
  concept to a clinical product aimed at improving patient care, and
  discusses lessons learned.

\end{abstract}

\section{Introduction}

New advances on research in applications of machine learning (ML) and
artificial intelligence in medicine are published on a regular
basis. However, there is lack of literature on translation of these
innovations to clinical practice, in particular for tests based on
molecular data. To fill the gap, we report on the development of
classifiers for detecting type and severity of infections in patients
who present in emergency departments (ED) with symptoms of acute
infection and sepsis, an unmet medical need [\citenum{XREF4}].

Current modalities for diagnosing infections mostly rely on detection
and identification of pathogens. However, this is inadequate because
in the majority of cases, pathogens are not found in blood or anywhere
else in the body~[\citenum{jain2015community}]. A more recent approach
relies on response of the immune system to the infection ({\em host
  response}). This could provide diagnostic and prognostic information
regardless of whether a pathogen is eventually identified.

From the ML perspective, diagnosing infections in the ED can be
reduced to two classification problems: infection type classification
(diagnosis) and illness severity classification (prognosis). The
classifiers use gene expression (abundance of mRNAs) of cells from
whole blood as input features. The diagnostic classifier estimates the
probability of the patient having bacterial, viral or no infection
(BVN), whereas the prognostic binary classifier estimates probability
of severe outcome in the given time window (SEV).

We used clinical adjudication~[\citenum{whitfield2024standardized}] as
the ground truth for the BVN classifier, and 30-day survival as the
ground truth for the SEV classifier. We used 29 mRNAs as input
numerical features, measured using a variety of measurement platforms
during training and validation, comprising both commercially-available
(microarrays, RNA-Seq and molecular barcoding technology
(NanoString$^{\text{\textregistered}}$), which are established
technologies for measuring gene expression) and our in-development
platform Myrna\textsuperscript{\texttrademark}, which uses a rapid
method called Loop-Mediated Isothermal Amplification (LAMP). The LAMP
experiments were performed using two approaches: 1. ``benchtop LAMP'',
which is not fully-automated and uses a commercially available
instrument. It was utilized before
TriVerity\textsuperscript{\texttrademark} became available, and
2. ``automated LAMP'' using TriVerity.

\begin{figure}[ht]
  \centering
  \includegraphics[width=0.25\textwidth]{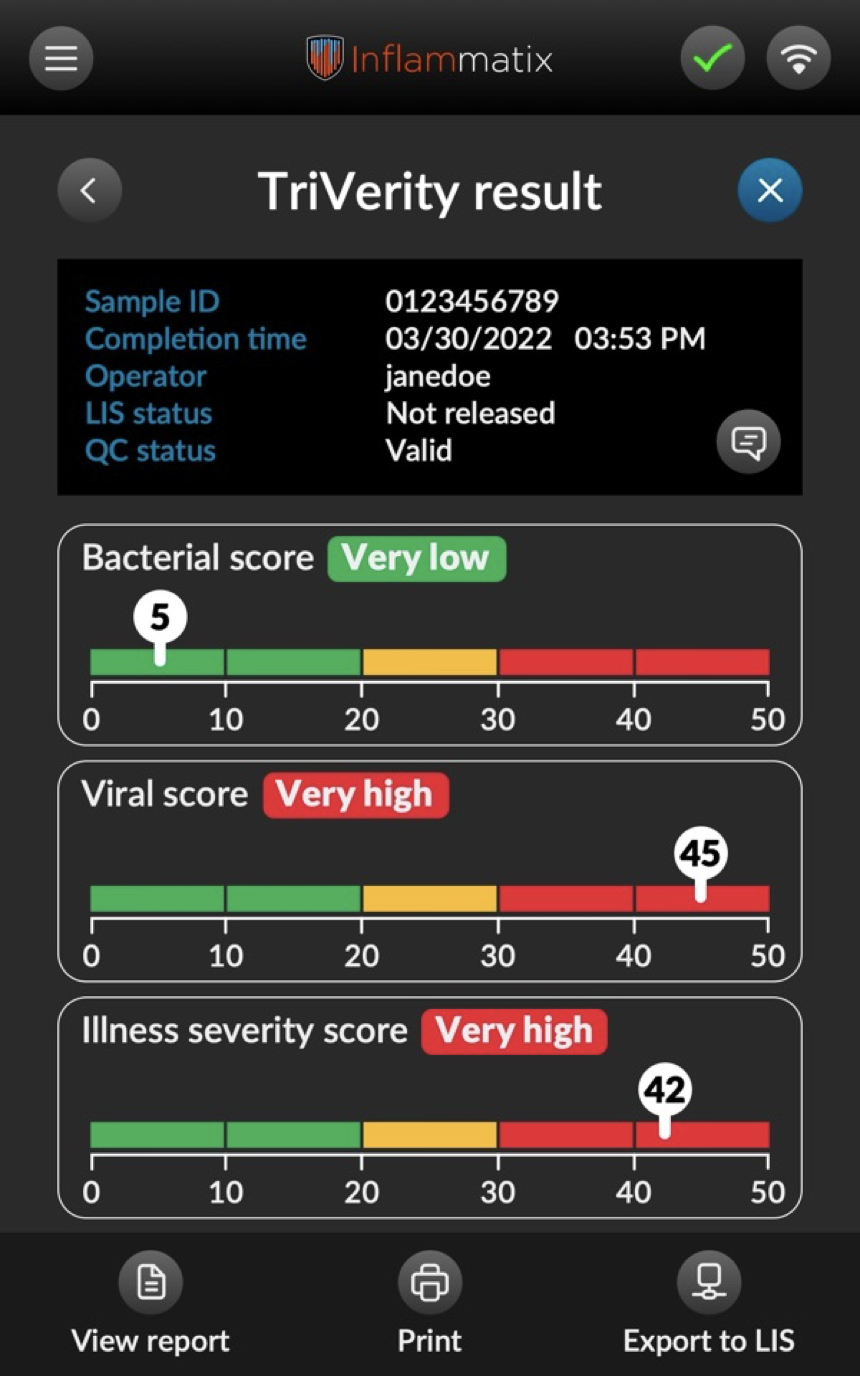}

  \caption{{\bf A sample TriVerity test report}. The scores are
    monotonic piecewise-linear functions of the probabilistic outputs
    of the BVN and SEV classifiers. During the development, severe
    illness was defined as death within 30 days of the test
    administration.}

  \label{fig:test_report}
\end{figure}

The final test produces three scores (Bacterial, Viral, Severity)
based on underlying classifier probabilities. The probabilities are
divided into five bands, corresponding to very low, low, moderate,
high, and very high probabilities of the corresponding class.  The
probabilities and bands are further transformed to scores on 0-50
scale and presented to the user (Fig~\ref{fig:test_report}). The test
report was developed using in-house expertise, feedback from ED and
clinical laboratory physicians, FDA predicate device clearances, and
additional market research.

The system software overview is described in
Section~\ref{SOFTWARE_ARCHITECTURE}.

\begin{figure}
  \centering
  \includegraphics[width=0.75\textwidth]{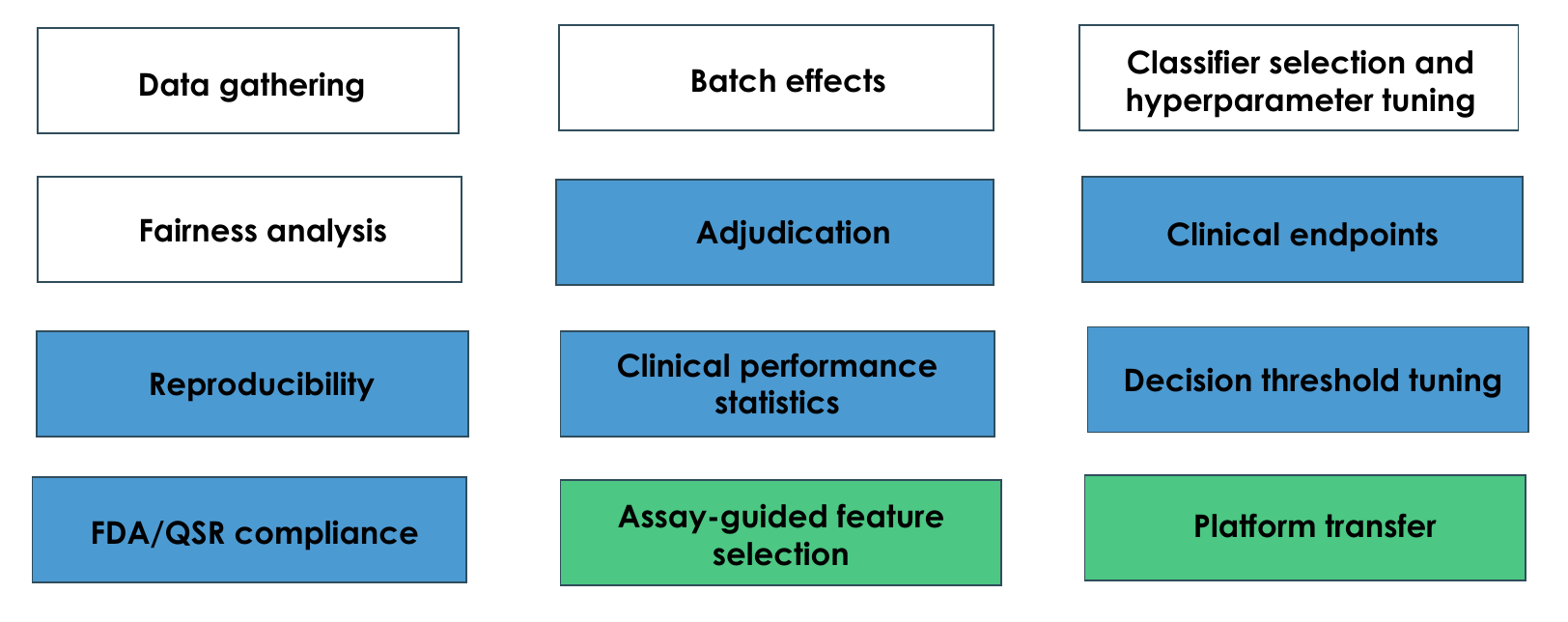}

  \caption{{\bf Key elements of the clinical classifier development
      process.} The white rectangles are generic machine learning
    activities. The blue rectangles are specific to clinical
    classifiers. The green rectangles are specific to clinical
    classifiers based on molecular features. QSR: Quality Systems
    Regulation; FDA: United States Food and Drug Administration. The
    names correspond to the manuscript sections, except {\em Clinical
      endpoints} which will be described elsewhere.}

  \label{fig:TriaMLElements}
\end{figure}

In this paper, we describe how the classifiers were developed and
their performance characteristics. Inspired
by~[\citenum{sculley2015hidden}], we show the key elements of the
classifier development in Fig.~\ref{fig:TriaMLElements}. As
in~[\citenum{sculley2015hidden}], we note that the classifier
selection and tuning represent a modest fraction of the overall effort
from academic prototype to product launch. The manuscript sections
correspond to the elements shown, except where noted. Given the
extensive validation of the 29-mRNA signature in academic prototypes
[\citenum{XREF1}, \citenum{XREF2}, \citenum{XREF3}], here we
specifically emphasize activities specific to commercially usable
point-of-care classifier development.

\section{Development data and classifier development}

First we describe the development data, with emphasis on properties
that have significant impact on the ML methods: batch effects and
platform transfer.

\subsection{Data gathering and adjudication}

The sources of development data were public (generated using
microarrays, from GEO~[\citenum{GEO}] and ArrayExpress~[\citenum{AE}])
and proprietary (generated using NanoString$^{\text{\textregistered}}$
and qPCR)~[\citenum{MAYHEW2020}]. The approach to selecting the
development data was guided by prior research on diagnosing infections
using a molecular approach [\citenum{XREF2}].

The development data was split into training and validation sets. The
training set was used for hyperparameter tuning with cross-validation
(CV), while the validation data was used for comparison of the tuned
classifiers, and never for classifier training nor hyperparameter
tuning (except exactly once to train the final locked
classifiers). This approach reduces bias compared with
CV~[\citenum{MAYHEW2020}].

The class labels ({\em clinical truth}) for the development data were
derived from the public records for public data and from a clinical
adjudication process for proprietary data. As such, the public data
were inherently more noisy than the proprietary data, but also better
representative of the heterogeneity in the real-world patient
population. The proprietary data were from company-sponsored clinical
studies, all with physician-adjudicated label assignment. The
proprietary studies were approved by institutional review boards.

The detailed information about the development datasets is given in
Section~\ref{DATASETS}.

\subsection{Batch effects}

The blood samples used to generate data for the development were
processed in different hospitals across independent studies and
generated using different technical platforms, on biologically and
clinically heterogenous patient populations (ED, intensive care units,
different races and ethnicities, different disease severity). We used
normalization to adjust different levels of gene expression, but the
normalization does not correct for technology- and study-induced batch
effects. To reduce their impact, we used validation and CV based on
{\em grouped} approach.

To reduce the platform-induced batch effects while maintaining
biological and clinical heterogeneity, we grouped the training and
validation data by gene expression platform. The training data were
almost exclusively ($\approx 98\%$) microarray data, whereas the
validation data were assayed on NanoString (during development) or the
target platform (for production classifiers).  The goal was to ensure
the robustness of the trained classifiers across a variety of
platforms. The advantage of this robustness was that it allowed use of
large amounts of publicly available datasets as they become available
to further improve accuracy of our classifiers, while maintaining our
ability to translate them to a point-of-care platform. The gap between
the training and validation performance served as an indicator of the
classifier's robustness with respect to the gene expression platform.

To reduce the study-induced batch effects, we used grouped
CV~[\citenum{groupedCV}]. The grouped CV was used in the
hyperparameter tuning to evaluate performance of candidate
hyperparameter configurations (HC).  In grouped CV, the folds were
configured such that different folds contained disjoint sets of
studies (in other words, all blood samples from a given study were
assigned to the same grouped CV fold). Analogous to the approach used
for platform-induced batch effects, HCs which exhibited poor
performance on the left-out folds (usually including study batch
effects) were automatically eliminated by the hyperparameter tuning
algorithms.

\subsection{Hyperparameter tuning, classifier selection and platform transfer}
\label{CF}

\begin{figure}
  \centering
  \includegraphics[width=1\textwidth]{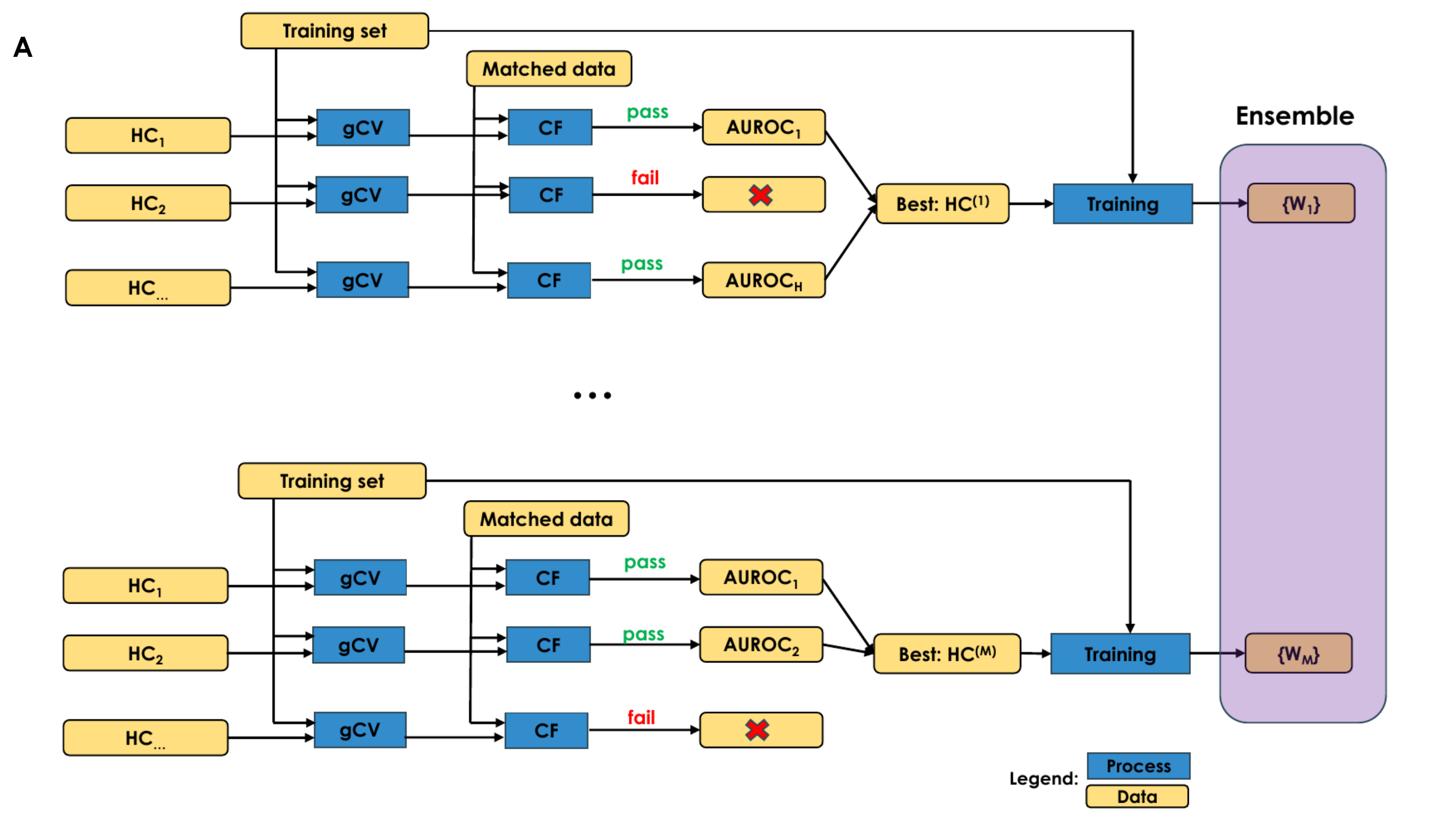}

  \caption{{\bf Overview of classifier tuning.} CF: Concordance
    Filtering (see Section~\ref{CF}); HC: Hyperparameter
    Configuration.}

  \label{fig:hc_tuning}
\end{figure}

An overview of hyperparameter tuning is shown in
Fig.~\ref{fig:hc_tuning}.

A key requirement of our classifiers was {\em platform transfer},
defined as generalization from the mostly public expression data,
which comprised the training set, to the target instrument. During
most of the classifier development, the target instrument was not
available because the device itself was being developed: the product
development took about six years from conception to the design freeze
(a timepoint at which device is considered finalized), whereas the
instrument data first became available about six months prior to the
design freeze. As a result, input features for the classifier training
were measured using different, commercially available instruments
(NanoString$^{\text{\textregistered}}$ was used as a proxy for
Myrna\textsuperscript{\texttrademark} during the development). The
platform transfer problem can be defined as ensuring that the
classifiers developed on the available platforms would perform
similarly on the target platform, once it becomes available. To enable
the platform transfer, we developed a {\em Concordance Filtering} (CF)
approach.

The platform transfer based on CF supports an efficient use of the
heterogenous data. It is an enabling step for applying the classifiers
developed on one or more platforms to a different or newly developed
platform. The number of samples available on our target platform was
relatively limited and deemed insufficient alone for training the
classifiers that would generalize to a global population. The goals of
the platform transfer were A) use all available data, on all
platforms, for classifier training, and B) maximize performance and
generalizability of the classifier on the target platform.

To develop an approach which satisfies the goals, we relied on the
following ideas:

\begin{itemize}

\item In early versions of the classifiers, optimize for adequate
  performance on the NanoString data as the highest-quality feature
  values available

\item Later, jointly optimize for high level of concordance (i.e.,
  similar gene expression values measured across replicate samples)
  between NanoString and Myrna

\item Thus, tuning the classifiers to achieve good concordance between
  platforms can be performed using relatively small unlabeled sets of
  samples assayed on both platforms

\end{itemize}

The CF algorithm removes, during hyperparameter tuning, the
classifiers which did not achieve adequate concordance between the
replicate samples. The concordance was measured by the coefficient of
determination ($R^2$) and/or Pearson correlation coefficient [PCC]
between NanoString$^{\text{\textregistered}}$- and
Myrna\textsuperscript{\texttrademark}-generated probabilities. In
other words, during the tuning, we applied a classifier to the set of
replicate NanoString/Myrna samples. If the concordance was below a
fixed minimal value, the classifier was removed from the tuning
process, regardless of clinical performance. For BVN, where there are
three classes, the CF filtering step was based on the minimum PCC or
$R^2$ for bacterial or viral class, which are of primary interest
because they are reported to the user. For SEV, the CF filtering was
based on the probability of the severe class.

Using this approach, we developed the instrument and classifiers in
parallel, thereby significantly reducing the product development
timelines and cost, and substantially boosting the number of samples
available for training. The final tuning of the classifiers using
Myrna data was done in the very last stages of the project and
required a modest number of samples. Specifically, for the last
versions of the classifiers we assayed a set of matching (replicate)
blood samples on NanoString and Myrna\textsuperscript{\texttrademark}
for use in the CF.

\subsection{Reproducibility}

Reproducibility refers to an important safety property of diagnostic
tests to generate clinically similar results using repeated
measurements from the same patient (for example, multiple tubes from a
single blood draw). It is conceptually similar to platform transfer
except that the replicates come from repeated measurements on the same
platform. Accordingly, we adopted the following approach: during CF,
add Myrna replicates to the set of NanoString/Myrna replicates. That
way, we ensured that the highly-ranked classifiers achieved high level
of reproducibility within and across platforms.

\subsection{Hyperparameter tuning: ensembling and loss}

Our approach used ensembling. We evaluated several variants of
ensembling to empirically find the best performing method. The
hyperparameter tuning was repeated $M = 10$ times independently, with
different random seeds and otherwise identical configurations. The
class probabilities were the averages of the $M$ output probabilities
generated by the ensemble members.

The loss function used for ranking the HCs was Area Under Receiver
Operating Characteristic (AUROC) for severity, and the multi-class
generalization of AUROC~[\citenum{HAND2001}] (mAUROC) for BVN
classifier. Ideally, the candidate classifiers should be ranked using
the performance statistics of the final product, such as sensitivity
and specificity. However, the clinical statistics
(Section~\ref{STATS}) involve setting of decision thresholds, which is
not amenable to full automation.

During CV, the mAUROC can be computed using averaging or pooling. In
averaging, mAUROC is estimated for each fold, and the final mAUROC is
the average across folds. In pooling, predicted probabilities for each
fold are recorded, and the final mAUROC is computed by pooling the
fold probabilities and estimating mAUROC using the pooled data. While
averaging may be preferred~[\citenum{POOLINGVSAVERAGING2007},
  \citenum{POOLINGVSAVERAGING2009}], in our case it was not an option
because the majority of studies did not have patients in all three
classes.

Based on expert judgment, we considered an increase in classification
performance of 0.01 AUROC units to be clinically significant.

We used Optuna~[\citenum{akiba2019optuna}] to select candidate
hyperparameter configurations (HC) from a search space of 1000. We
chose a fixed value of five folds for all CV experiments.


\subsection{Assay-guided feature selection}

Academic research identified 29 candidate genes (a {\em signature})
for diagnosis and prognosis of presence, type and severity of
infectious diseases [\citenum{XREF1}, \citenum{XREF2},
  \citenum{XREF3}]. They formed an initial set for the search for
genes compatible with the target device. The search is described in
Section~\ref{FEATURE_SELECTION}.

\subsection{Classifier lock}

The classifier lock is the setting of classifiers' parameters
(weights) to final optimized values for deployment in clinical care. A
key practical decision was to choose between the following options:

\begin{enumerate}

\item Pool the training and validation data and train the final
  classifier using the pooled dataset with hyperparameters found by
  hyperparameter tuning. In this case, training data is maximized, and
  includes the target platform data, but there is no data left to
  validate the locked classifiers.

\item Lock the models trained on training data. The classifiers are
  validated using the validation set. However, the training data is
  reduced, and does not use the target platform.

\end{enumerate}

We opted for pooling, based on the strong preference to incorporate
the target platform in the training of the locked classifiers.

\subsection{Fairness analysis}

The performance of the development versions of the classifiers across
race, age and sex subgroups was reported
in~[\citenum{mayhew2022towards}]. The performance of the production
classifiers will be published in a future manuscript, following the
completion of the pivotal trials.

\subsection{QSR/FDA compliance}

The Myrna\textsuperscript{\texttrademark} system and
TriVerity\textsuperscript{\texttrademark} Test, including the
classifiers, are subject to the Quality Systems Regulation by the
United States FDA. The regulation is codified in Title 21 of the Code
of Federal Regulations (CFR) Part 820~[\citenum{QSR}] and is meant to
ensure quality and safety of medical devices intended for human
use. It is a complex topic outside of scope of this
manuscript. Machine learning activities impacted by the Part 820
include documentation, personnel training, corrective and preventive
actions, hazard analysis, statistical techniques and classifier
updates.

\section{Clinical performance statistics}
\label{STATS}

The key test performance characteristics comprise likelihood ratios
for individual test bands, band sensitivities and specificities,
coverage and monotonicity.

For tests with two possible test results, positive and negative
(binary tests), the positive diagnostic likelihood ratio (LR) is
defined as the ratio of the probability of a positive test result in
patients with the disease to the probability of a positive test result
in patients without the disease. The negative diagnostic ratio is
defined as the ratio of the probability of a negative test result in
patients with the disease to the probability of a negative test result
in patients without the disease~[\citenum{HAYDEN}]. The two LRs are
the properties of a classifier and a decision threshold, and fully
characterize the test performance. However, TriVerity has five
possible test results (assigned bands,
Fig~\ref{fig:test_report}). This impacts the definitions and
interpretations of the standard performance statistics. The definition
of the LRs in such cases is non-obvious because it is not apparent
what constitutes positive and negative test results for the bands. For
computing the LRs for band $i$, we considered all samples assigned by
the classifiers to band $i$ as {\em predicted positive}, and samples
assigned outside the band as {\em predicted negative}. This permits an
intuitive behavior of monotonically increasing likelihood ratios from
band 1 to band 5 (an established desirable property of multi-band
tests~[\citenum{RAPID}]). Given the above convention, the positive LRs
become:

\begin{align}
  LR_i &= \frac{\frac{x_i}{X}}{\frac{y_i}{Y}}, i \in 1..5, X = \sum_{i=1}^5 x_i, Y = \sum_{i=1}^5 y_i
\end{align}

where $x_i, y_i$ are the number of samples with ground truth {\bf POS}
and {\bf NEG} in band $i$, respectively.

\section{Decision threshold tuning}

Decision thresholds help clinicians make diagnosis and treatment
decisions. Each TriVerity output probability is split in five bands using
four thresholds (Fig~\ref{fig:test_report}).

The decision threshold tuning required the specification of the
threshold-tuning-set (TTS) and a decision tuning algorithm. The
assumption is that the samples in TTS have been analyzed using a
classifier, producing predictive probabilities. Subsequently, we tune
decision thresholds using the TTS predicted probabilities and the
combination of decision tuning algorithm and manual finetuning.

In problems with two classes and a binary test report, the threshold
tuning is typically achieved by finding a clinically best trade-off
between sensitivity and specificity, for example by expert
judgment. However, TriVerity\textsuperscript{\texttrademark} requires
setting of 8 thresholds for BVN and 4 for SEV
(Fig~\ref{fig:test_report}). To facilitate the threshold selection
during model tuning, we introduced an algorithm called GAT (Genetic
Algorithm for Thresholds, [\citenum{GAT}]). GAT is based on genetic
(evolutionary) optimization as implemented in the DEAP python
library~[\citenum{DEAP}]. Each individual ({\em chromosome}) in GAT is
a set of the decision threshold values. GAT operates by maximizing a
fitness function through evolutionary computation until the chromosome
with the best fitness is found. GAT is applied to each test output
(bacterial, viral, severity) independently. It is defined as follows:

{\bf Input}: desired (target) values for LR1, LR5, coverage, and
percent moderate

{\bf Output}: best thresholds (4 values)

\begin{enumerate}

\item The initial population for the evolutionary (genetic) algorithm
  is randomly generated. A set of chromosomes, each representing a
  potential solution to the problem, is created. The chromosome
  corresponding to a solution has four values, representing the four
  thresholds needed to split the probabilities into five bands.

\item The fitness of each chromosome in the population is evaluated
  using a fitness function. The function assigns a fitness score to
  each chromosome based on how well it fits the desired criteria for
  LR1, LR5 and coverage. The coverage is only considered if the value
  is below the target value, so that coverage exceeding the target is
  not penalized.

\item A new generation is created by selecting parents according to
  their fitness. Offspring are created using crossover and mutation
  operations. The individuals with the top 20\% fitness are always
  kept in the population.

\item Steps 2 and 3 are repeated for a given number of iterations.

\end{enumerate}

The specification of TTS is, to the best of our knowledge, an unsolved
problem. The problem is that, following classifier development, a
separate dataset may be required to tune the decision thresholds, and
that may not always be available. We considered the following options:
A) using the same training data for threshold tuning and performance
evaluation and B) setting aside a separate set of data exclusively for
TTS. The approach A is a resubstitution approach, which may introduce
optimistic bias. The approach B may not use the valuable data
efficiently. We decided between the two approaches as follows:

\begin{itemize}

\item Tune the thresholds using resubstitution and GAT algorithm. This
  produces thresholds $T$

\item Estimate clinical performance $R$ using TTS probabilities and
  thresholds $T$. $R$ is a set of relevant statistics such as LRs and
  coverage

\item Tune the thresholds using a reduced-bias approach (the reduction
  refers to bias in comparison with the resubstitution approach). This
  means using different set of data for threshold tuning
  vs. classifier performance assessment

\item Record the resulting clinical performance $U$

\item If the difference between $R$ and $U$ is clinically
  insignificant, use the data-efficient resubstitution approach for
  decision thresholds tuning.

\end{itemize}

The algorithm requires reduced-bias threshold selection. To that end,
we introduced the following algorithm for decision threshold CV:

\begin{itemize}

\item Split TTS (i.e., probabilities) into $K$ folds. Each fold
  consists of learning set (LSET) and cross-validation set (CVSET)

\item For each (LSET, CVSET) pair:

  \begin{itemize}

  \item Tune decision thresholds using GAT algorithm and LSET dataset

  \item Assign the CVSET probabilities to different bands using the
    tuned thresholds

  \end{itemize}

\item Compute the TTS clinical performance using the pooled band
  assignments

\end{itemize}

The threshold CV computes reduced-bias estimate of the clinical
performance when thresholds are tuned using independent data. This
estimate is then used to decide whether the data-efficient
resubstitution method has acceptable bias. If the conclusion is
positive, we can use the same dataset for threshold tuning and
estimation of performance; otherwise, a separate TTS is needed.

The selection of thresholds during classifier development used GAT due
to good performance, automation and high throughput; however, for the
final classifier lock, the thresholds were manually finetuned, to
achieve the best trade-off between different clinical performance
statistics.

\section{Results}

The quality of input data and normalization approaches were critical
to the development. The BVN mAUROC improved from under 0.7 in the
early days of the development, which was not viable for launching the
product, to 0.83, which provides favorable overall clinical
performance. We hypothesize that the progress was achieved by
increasing the quality and quantity of the public and proprietary
data, and improving development methodology. The data increase
qualitatively changed the type of top-ranked classifiers, which
oscillated between multi-layer perceptron (MLP) and LOGR, until the
introduction of platform transfer. At design freeze, the BVN
development data comprised 4306 training and 679 validation samples,
derived from 47 and 7 studies, respectively. The SEV development data
comprised 2758 training and 723 validation samples, derived from 38
and 8 studies, respectively.

We periodically evaluated seven leading classification algorithms at
different time points during the development. A late-stage snapshot is
shown in Table~\ref{tab:classifier_comparison}. The winning classifier
was LOGR for both BVN and SEV problems.

\begin{table}
  \caption{{\bf Classifier accuracy and platform-to-platform
      concordance comparison.} MLP: Multi-layer Perceptron; LOGR:
    Logistic Regression; RBF: Support Vector Machine with Radial Basis
    Function kernel; SVM: Support Vector Machine with linear
    kernel; LGBM: Light GBM classifier; XGB: XGBoost; GPC: Gaussian
    Process Classifier. Concordance: Pearson correlation coefficient
    between replicate samples assayed on
    NanoString$^{\text{\textregistered}}$ and LAMP. The classifiers
    are sorted by validation mAUROC. The ties are resolved by
    concordance.}
  \label{tab:classifier_comparison}
  \centering
  \begin{tabular}{cccc}
    \toprule
    Algorithm    & CV mAUROC [95\% CI] & Val. mAUROC [95\% CI]  & Concordance [95\% CI] \\
    \midrule
    MLP          & 0.88 [0.88, 0.89]   & 0.84 [0.81, 0.86]      & 0.94 [0.93, 0.94]     \\
    LOGR         & 0.89 [0.88, 0.89]   & 0.84 [0.81, 0.86]      & 0.93 [0.92, 0.94]     \\
    RBF          & 0.87 [0.86, 0.88]   & 0.83 [0.80, 0.86]      & 0.95 [0.94, 0.96]     \\
    SVM          & 0.87 [0.86, 0.87]   & 0.83 [0.80, 0.86]      & 0.92 [0.91, 0.93]     \\
    LGBM         & 0.87 [0.86, 0.88]   & 0.83 [0.80, 0.86]      & 0.88 [0.86, 0.89]     \\
    XGB          & 0.87 [0.86, 0.88]   & 0.82 [0.78, 0.84]      & 0.80 [0.78, 0.82]     \\
    GPC          & 0.86 [0.85, 0.87]   & 0.81 [0.77, 0.84]      & 0.76 [0.73, 0.78]     \\
    \bottomrule
  \end{tabular}
\end{table}

The greedy forward search did not yield a subset of genes which
improved clinical or analytical performance for at least one of the
classifiers (results not shown). Based on this observation, we decided
to finalize the product using 29 genes.

The final results of TriVerity\textsuperscript{\texttrademark} classifiers
are shown in Table~\ref{tab:clinical_performance}. Importantly, we
found that for $N = 534$ patients for which both BVN and SEV ground
truths were available, 73\% were assigned by the classifiers to at
least one of the outer bands (1 or 5). This is clinically important
because those diagnoses provide the most confidence in their
result. In contrast, only 0.2\% of the patients in the common set were
assigned to band 3 for all three scores. This is important because the
band 3 has LRs that are clinically uniformative.

\begin{table}
  \caption{{\bf Clinical performance of the winning
      TriVerity\textsuperscript{\texttrademark} classifiers.} All samples
    were processed using Myrna\textsuperscript{\texttrademark}
    Instrument. PPA: Positive Percent Agreement (equivalent to
    sensitivity when ground truth is not 100\% reliable); NPA:
    Negative Percent Agreement (equivalent to specificity when ground
    truth is not 100\% reliable). Coverage is the percent of patients
    assigned to the indicated band(s). {\em Moderate} refers to band
    3.}
  \label{tab:clinical_performance}
  \centering
  \begin{tabular}{cccc}
    \toprule
    Band(s)      & Statistic           & Training Value [95\% CI]  & Val. Value [95\% CI]    \\
    \midrule
                 & BVN mAUROC          & 0.886 [0.877, 0.894]      & 0.833 [0.806, 0.861]    \\
    1            & Bacterial LR        & 0.09  [0.07,  0.11 ]      & 0.08  [0.03,  0.14 ]    \\
    5            & Bacterial LR        & 7.1   [6.2,   8.2  ]      & 7.6   [4.8,   15   ]    \\
    1            & Bacterial PPA       & 95    [94,    97   ]      & 98    [96,    99   ]    \\
    5            & Bacterial NPA       & 93    [92,    94   ]      & 95    [93,    98   ]    \\
    1, 5         & Bacterial coverage  & 60    [58,    61   ]      & 35    [31,    39   ]    \\
    3            & Bacterial moderate  & 7.1   [6.3,   7.8  ]      & 16    [13,    19   ]    \\
    1            & Viral LR            & 0.07  [0.05,  0.10 ]      & 0.13  [0.05,  0.22 ]    \\
    5            & Viral LR            & 8.1   [7.3,   9.1  ]      & 9.6   [7.0,   14   ]    \\
    1            & Viral PPA           & 97    [96,    98   ]      & 95    [92,    98   ]    \\
    5            & Viral NPA           & 90    [89,    91   ]      & 93    [91,    95   ]    \\
    1, 5         & Viral coverage      & 54    [53,    56   ]      & 52    [48,    55   ]    \\
    3            & Viral moderate      & 13    [12,    14   ]      & 15    [12,    18   ]    \\
    1, 5         & BV coverage         & 72    [71,    74   ]      & 58    [54,    62   ]    \\
    \midrule
                 & SEV AUROC           & 0.886 [0.866, 0.906]      & 0.793 [0.738, 0.843]    \\
    1            & SEV LR              & 0.04  [0.01,  0.09 ]      & 0.16  [0.04,  0.33 ]    \\
    5            & SEV LR              & 7.3   [5.8,   9.0  ]      & 8.9   [5.2,   16   ]    \\
    1            & SEV PPA             & 98    [95,    100  ]      & 94    [89,    99   ]    \\
    5            & SEV NPA             & 93    [92,    94   ]      & 97    [95,    98   ]    \\
    5            & SEV coverage        & 61    [59,    63   ]      & 40    [36,    44   ]    \\
    3            & SEV moderate        & 6.5   [5.5,   7.4  ]      & 15    [12,    17   ]    \\
    \bottomrule
  \end{tabular}
\end{table}

We observed 0.886 BVN mAUROC in training and 0.833 in validation. We
hypothesize that the principal causes of the gap are the major
platform difference (BVN training is almost entirely microarray data,
whereas validation is Myrna\textsuperscript{\texttrademark} data) and
significant differences in patient populations.

We observed 0.886 SEV AUROC in training and 0.793 in validation.
Besides platform gap, the severity gap was due to inclusion of
pediatric patients in severity training set. The rationale for this
design choice is that removal of pediatric patients from training had
limited to no impact on the validation results, and it may improve the
performance in a future expanded indication for the test.

To estimate the impact of CF, we performed an ablation study in
Section~\ref{CF_ABLATION}.

\section{Discussion}

We performed successful transfer of a research prototype of a clinical
infectious disease diagnostic test to a design freeze product ready
for independent verification and validation, across heterogenous data
environment. The results suggest that the BVN and SEV ML classifiers
perform adequately for clinical
deployment~[\citenum{DUCHARME2020}]. Approximately three out of four
patients are likely to receive at least one clinically highly
informative score (bands 1 or 5), and almost none (0.2\% in the
validation set) are likely to receive an uninformative test report.
Based on our assessment of the clinical utility and commercial
potential of the validation results, we built and locked the
production classifiers and initiated pivotal trials.

We found that concordance filtering significantly improved the quality
of platform transfer for some classification algorithms (MLP). For
others, like LOGR, it served as assurance that the winning classifier
would meet the platform transfer requirements.

The validation performance showed that grouped CV effectively
countered the platform and study batch effects. While the effects were
not completely eliminated, they were nevertheless sufficiently
attenuated to enable clinical application of the system.

The classifier comparison showed relatively small but clinically
relevant differences in performance by AUROC. The gradient-boosting
classifier (XGBoost and Light GBM), generally considered top choice
for classification of tabular data, underperformed in our use case,
both by classification performance and concordance among
platforms. This is a somewhat surprising finding, potentially caused
by strictly numeric type of the input features. We also observed
consistent underperformance by Gaussian Process Classifier, which may
be useful learning because this classifier is relatively rarely
evaluated in ML manuscripts. These findings are under active
investigations.

Arguably, our use of pooling the predicted probabilities across CV
folds is biased. However, the clinical application and the composition
of development data necessitated this approach. We found the resulting
performance to be robust in validation set and across gene expression
platforms.

Our work and manuscript have limitations. We presented development
data; a prospective evaluation of the clinical validity is currently
in progress. The clinical utility of
TriVerity\textsuperscript{\texttrademark} will be determined in
prospective and randomized trials, currently in planning stages. The
adjudicated ground truth for TriVerity development is imperfect and
may affect the performance statistics reported here.

\section{Conclusion}

We presented the development of clinical classifiers for rapid
diagnosis and prognosis of acute infections and sepsis in emergency
departments based on blood draw. Internal evaluations suggest that the
product is adequate for clinical use and commercial launch, pending
clearance by the United States FDA.

\bibliographystyle{IEEEtran} 
\bibliography{Tria} 

\section{Technical Appendices}

\subsection{Software architecture}
\label{SOFTWARE_ARCHITECTURE}

\begin{figure}
  \centering
  \includegraphics[width=1\textwidth]{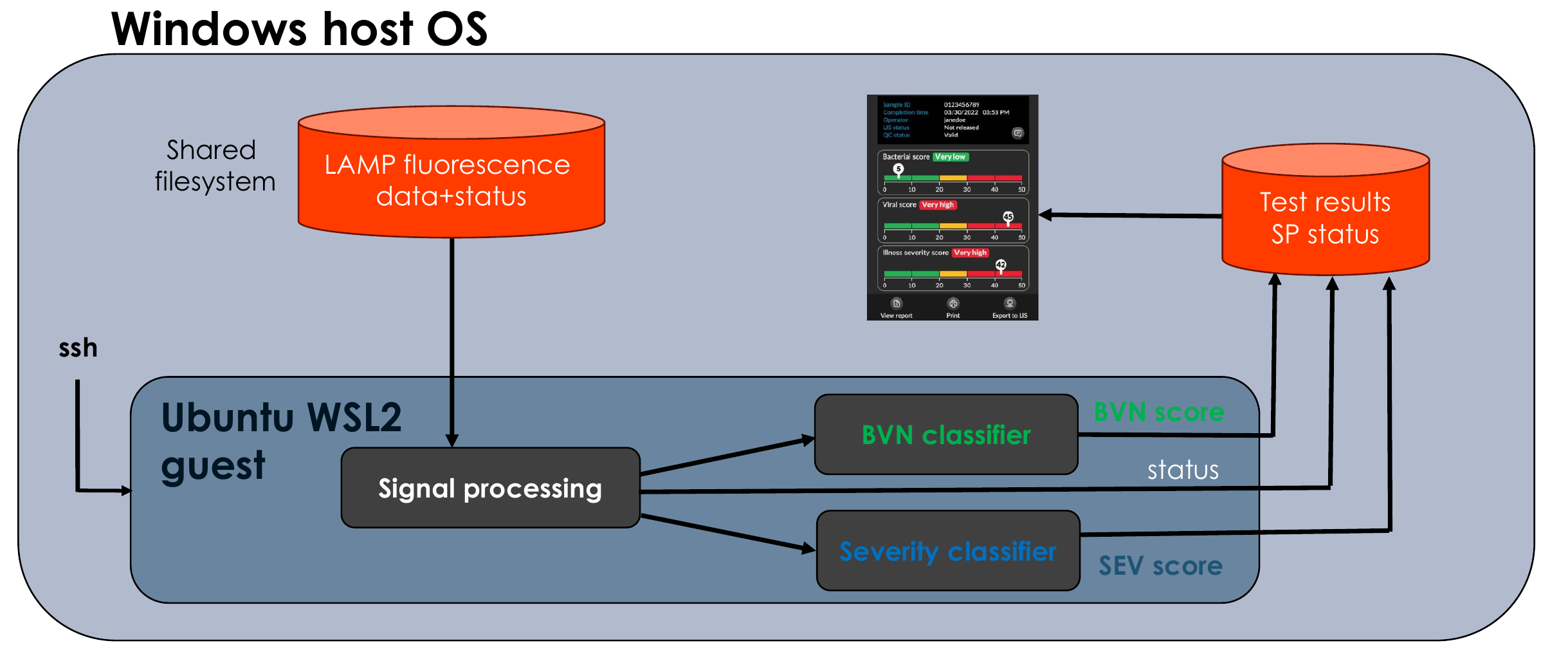}

  \caption{{\bf The system software overview.} BVN:
    Bacterial/Viral/Noninfected; SEV: Severity; SP: Signal Processing;
    LAMP: Loop-Mediated Isothermal Amplification.}

  \label{fig:RP_diagram}
\end{figure}

The Myrna\textsuperscript{\texttrademark} device runs Windows OS. The
software architecture, including classifiers, is shown in
Fig.~\ref{fig:RP_diagram}.

\subsection{Datasets for training and validation}
\label{DATASETS}

Development data are summarized in Tables~\ref{tab:bvn_study_summary}
and ~\ref{tab:sev_study_summary}. Healthy controls are not listed in
the tables because the detailed demographic information was not
available.
    
\begin{table}
  \centering

  \caption{{\bf Development datasets for diagnostic classifier
      (bacterial, viral, noninfected).} ICU: Intensive Care Unit. IQR:
    Interquartile Range. Unk.: Unknown. COPD: Chronic Obstructive
    Pulmonary Disease. CAP: Community-Acquired Pneumonia. SIRS:
    Systemic Inflammatory Response Syndrome.}

  \resizebox{\textwidth}{!}{\begin{tabular}{|c|c|c|c|c|c|c|c|c|c|c|c|}
      \hline
      Study & Author & Description & N & Age$^{*}$ & Male$^{*}$ & Platform & Country & Bacterial & Viral & Non-infected \\ \hline
      \href{http://www.ebi.ac.uk/arrayexpress/experiments/E-MEXP-3589/}{EMEXP3589} & Almansa & \shortstack{ Patients hospitalized with COPD \\ exacerbation } & 27 & Unk. & 16 (59) & Agilent & Spain & 4 (15) & 5 (19) & 18 (67) \\ \hline
      \href{http://www.ebi.ac.uk/arrayexpress/experiments/E-MTAB-1548/}{EMTAB1548} & Almansa & \shortstack{ Surgical patients with sepsis \\ (EXPRESS) } & 155 & 72.0 (IQR 61.0-78.0) & 95 (61) & Agilent & Spain & 82 (53) & 0 (0) & 73 (47) \\ \hline
      \href{https://www.ebi.ac.uk/biostudies/arrayexpress/studies/E-MTAB-3162}{EMTAB3162} & van den Ham & Patients with dengue & 21 & 20.0 (IQR 17.0-28.0) & 11 (52) & Affymetrix & Indonesia & 0 (0) & 21 (100) & 0 (0) \\ \hline
      \href{http://www.ebi.ac.uk/arrayexpress/experiments/E-MTAB-5273/}{EMTAB5273} & Burnham & \shortstack{ Sepsis due to faecal \\ peritonitis or pneumonia } & 129 & 66.0 (IQR 51.0-76.0) & 59 (46) & Illumina & United Kingdom & 119 (92) & 0 (0) & 10 (8) \\ \hline
      \href{http://www.ebi.ac.uk/arrayexpress/experiments/E-MTAB-5273/}{EMTAB5274} & Burnham & \shortstack{ Sepsis due to faecal \\ peritonitis or pneumonia } & 108 & 71.0 (IQR 62.75-77.0) & 69 (64) & Illumina & United Kingdom & 108 (100) & 0 (0) & 0 (0) \\ \hline
      \href{https://www.ebi.ac.uk/biostudies/arrayexpress/studies/E-MTAB-5638}{EMTAB5638} & Almansa & Ventilator-associated pneumonia in ICU & 17 & Unk. & 0 (0) & Agilent & Spain & 0 (0) & 0 (0) & 17 (100) \\ \hline
      \href{https://www.ncbi.nlm.nih.gov/geo/query/acc.cgi?acc=GSE42026}{GSE42026} & Herberg & \shortstack{ Severe H1N1/09 influenza or \\ bacterial infection } & 92 & 1.25 (IQR 0.38-4.0) & 33 (36) & Illumina & United Kingdom & 18 (20) & 41 (45) & 33 (36) \\ \hline
      \href{https://www.ncbi.nlm.nih.gov/geo/query/acc.cgi?acc=GSE42834}{GSE42834} & Bloom & Bacterial pneumonia or sarcoidosis & 200 & Unk. & 32 (16) & Illumina & United Kingdom & 14 (7) & 0 (0) & 186 (93) \\ \hline
      \href{https://www.ncbi.nlm.nih.gov/geo/query/acc.cgi?acc=GSE47655}{GSE47655} & Stone & Severe anaphylaxis in ED & 12 & Unk. & 0 (0) & Affymetrix & Australia & 0 (0) & 0 (0) & 12 (100) \\ \hline
      \href{https://www.ncbi.nlm.nih.gov/geo/query/acc.cgi?acc=GSE51808}{GSE51808} & Kwissa & Patients with dengue & 37 & Unk. & 0 (0) & Affymetrix & Thailand & 0 (0) & 28 (76) & 9 (24) \\ \hline
      \href{https://www.ncbi.nlm.nih.gov/geo/query/acc.cgi?acc=GSE57065}{GSE57065} & Cazalis & Septic shock & 53 & 62.0 (IQR 54.25-76.0) & 19 (36) & Affymetrix & France & 28 (53) & 0 (0) & 25 (47) \\ \hline
      \href{https://www.ncbi.nlm.nih.gov/geo/query/acc.cgi?acc=gse57183}{GSE57183} & Shenoi & Febrile children with SJIA & 14 & 3.6 (IQR 3.3-7.3) & 5 (36) & Illumina & United States of America & 0 (0) & 0 (0) & 14 (100) \\ \hline
      \href{https://www.ncbi.nlm.nih.gov/geo/query/acc.cgi?acc=GSE25504}{GSE25504} & Smith & Neonatal sepsis & 39 & 0.0 (IQR 0.0-0.0) & 19 (49) & Affymetrix & United Kingdom & 29 (74) & 4 (10) & 6 (15) \\ \hline
      \href{https://www.ncbi.nlm.nih.gov/geo/query/acc.cgi?acc=GSE60244}{GSE60244} & Suarez & Lower respiratory tract infections & 133 & 63.0 (IQR 50.0-77.0) & 37 (28) & Illumina & United States of America & 22 (17) & 71 (53) & 40 (30) \\ \hline
      \href{https://www.ncbi.nlm.nih.gov/geo/query/acc.cgi?acc=gse61821}{GSE61821} & Hoang & \shortstack{ Influenza patients of varying \\ severity } & 48 & 40.0 (IQR 19.75-51.0) & 24 (50) & Illumina & Vietnam & 0 (0) & 48 (100) & 0 (0) \\ \hline
      \href{https://www.ncbi.nlm.nih.gov/geo/query/acc.cgi?acc=gse63881}{GSE63881} & Hoang & \shortstack{ Pediatric patients with Kawasaki \\ disease } & 171 & 2.75 (IQR 1.42-4.25) & 102 (60) & Illumina & United States of America & 0 (0) & 0 (0) & 171 (100) \\ \hline
      \href{https://www.ncbi.nlm.nih.gov/geo/query/acc.cgi?acc=GSE64456}{GSE64456} & Mahajan & \shortstack{ Febrile infants with bacterial \\ or viral infection in ED } & 219 & 0.1 (IQR 0.06-0.13) & 106 (48) & Illumina & United States of America & 89 (41) & 111 (51) & 19 (9) \\ \hline
      \href{https://www.ncbi.nlm.nih.gov/geo/query/acc.cgi?acc=GSE65682}{GSE65682} & Scicluna & \shortstack{ Suspected but negative for \\ CAP } & 75 & 59.0 (IQR 48.0-67.0) & 22 (28) & Affymetrix & Netherlands & 0 (0) & 0 (0) & 75 (100) \\ \hline
      \href{https://www.ncbi.nlm.nih.gov/geo/query/acc.cgi?acc=GSE66099}{GSE66099} & Sweeney & \shortstack{ Pediatric ICU patients with \\ sepsis, SIRS } & 197 & 2.45 (IQR 1.0-5.88) & 94 (48) & Affymetrix & United States of America & 109 (55) & 11 (6) & 77 (39) \\ \hline
      \href{https://www.ncbi.nlm.nih.gov/geo/query/acc.cgi?acc=GSE67059}{GSE67059} & Heinonen & Children with rhinovirus & 101 & 0.83 (IQR 0.3-1.29) & 53 (52) & Illumina & United States of America & 0 (0) & 80 (79) & 21 (21) \\ \hline
      \href{}{glueBuffyHCSS} & Multiple authors & \shortstack{ Trauma patients with or \\ without infection } & 372 & 32.0 (IQR 24.0-41.0) & 98 (26) & Affymetrix & United States of America & 47 (13) & 0 (0) & 325 (87) \\ \hline
      \href{https://www.ncbi.nlm.nih.gov/geo/query/acc.cgi?acc=GSE68310}{GSE68310} & Zhai & \shortstack{ Outpatients with acute respiratory \\ viral infections } & 347 & 20.96 (IQR 20.09-22.76) & 50 (14) & Illumina & United States of America & 0 (0) & 104 (30) & 243 (70) \\ \hline
      \href{https://www.ncbi.nlm.nih.gov/geo/query/acc.cgi?acc=GSE69528}{GSE69528} & Khaenam & \shortstack{ Sepsis, many cases from \\ burkholderia } & 138 & Unk. & 0 (0) & Illumina & Thailand & 83 (60) & 0 (0) & 55 (40) \\ \hline
      \href{https://www.ncbi.nlm.nih.gov/geo/query/acc.cgi?acc=GSE72810}{GSE72810} & Herberg & \shortstack{ Pediatric patients with bacterial \\ or viral infection } & 15 & 1.83 (IQR 0.88-3.29) & 8 (53) & Illumina & United Kingdom & 5 (33) & 10 (67) & 0 (0) \\ \hline
      \href{https://www.ncbi.nlm.nih.gov/geo/query/acc.cgi?acc=GSE73461}{GSE73461} & Wright & \shortstack{ Pediatric patients with bacterial \\ or viral infection or other inflammatory diseases } & 363 & 2.79 (IQR 0.92-8.81) & 165 (45) & Illumina & United Kingdom & 52 (14) & 94 (26) & 217 (60) \\ \hline
      \href{https://www.ncbi.nlm.nih.gov/geo/query/acc.cgi?acc=gse77087}{GSE77087} & de Steenhuijsen Piters & \shortstack{ Infants with respiratory synctial \\ virus } & 59 & 0.45 (IQR 0.14-0.69) & 25 (42) & Illumina & United States of America & 0 (0) & 41 (69) & 18 (31) \\ \hline
      \href{https://www.ncbi.nlm.nih.gov/geo/query/acc.cgi?acc=GSE77791}{GSE77791} & Plassais & Severe burn shock & 30 & 48.0 (IQR 40.25-55.0) & 21 (70) & Affymetrix & France & 0 (0) & 0 (0) & 30 (100) \\ \hline
      \href{https://www.ncbi.nlm.nih.gov/geo/query/acc.cgi?acc=GSE82050}{GSE82050} & Tang & \shortstack{ Moderate and severe influenza \\ infection } & 39 & 64.5 (IQR 48.5-74.25) & 13 (33) & Agilent & Germany & 0 (0) & 24 (62) & 15 (38) \\ \hline
      \href{https://www.ncbi.nlm.nih.gov/geo/query/acc.cgi?acc=gse103842}{GSE103842} & Rodriguez-Fernandez & \shortstack{ Infants with respiratory synctial \\ virus } & 74 & 0.25 (IQR 0.17-0.44) & 39 (53) & Illumina & United States of America & 0 (0) & 62 (84) & 12 (16) \\ \hline
      \href{https://www.ncbi.nlm.nih.gov/geo/query/acc.cgi?acc=GSE111368}{GSE111368} & Dunning & Influenza H1N1 and B & 33 & 38.0 (IQR 29.0-49.0) & 15 (45) & Illumina & United Kingdom & 0 (0) & 33 (100) & 0 (0) \\ \hline
      \href{https://www.ncbi.nlm.nih.gov/geo/query/acc.cgi?acc=GSE13015}{GSE13015gpl6102} & Pankla & \shortstack{ Sepsis, many cases from \\ burkholderia } & 55 & 54.0 (IQR 48.0-61.0) & 26 (47) & Illumina & Thailand & 45 (82) & 0 (0) & 10 (18) \\ \hline
      \href{https://www.ncbi.nlm.nih.gov/geo/query/acc.cgi?acc=GSE13015}{GSE13015gpl6947} & Pankla & \shortstack{ Sepsis, many cases from \\ burkholderia } & 20 & 49.0 (IQR 43.5-59.5) & 6 (30) & Illumina & Thailand & 15 (75) & 0 (0) & 5 (25) \\ \hline
      \href{https://www.ncbi.nlm.nih.gov/geo/query/acc.cgi?acc=GSE21802}{GSE21802} & Bermejo-Martin & Pandemic H1N1 in ICU & 14 & Unk. & 0 (0) & Illumina & Unk. & 0 (0) & 10 (71) & 4 (28) \\ \hline
      \href{https://www.ncbi.nlm.nih.gov/geo/query/acc.cgi?acc=GSE22098}{GSE22098} & Berry & \shortstack{ Bacterial infection or other \\ inflammatory conditions } & 274 & 16.0 (IQR 11.0-26.0) & 59 (22) & Illumina & Unk. & 52 (19) & 0 (0) & 222 (81) \\ \hline
      \href{https://www.ncbi.nlm.nih.gov/geo/query/acc.cgi?acc=GSE27131}{GSE27131} & Berdal & Severe H1N1 & 14 & 39.0 (IQR 35.75-44.75) & 10 (71) & Affymetrix & Norway & 0 (0) & 7 (50) & 7 (50) \\ \hline
      \href{https://www.ncbi.nlm.nih.gov/geo/query/acc.cgi?acc=GSE28750}{GSE28750} & Sutherland & Sepsis or post-surgical SIRS & 21 & Unk. & 0 (0) & Affymetrix & Australia & 10 (48) & 0 (0) & 11 (52) \\ \hline
      \href{https://www.ncbi.nlm.nih.gov/geo/query/acc.cgi?acc=gse28991}{GSE28991} & Naim & Patients with suspected dengue & 11 & Unk. & 0 (0) & Illumina & Unk. & 0 (0) & 11 (100) & 0 (0) \\ \hline
      \href{https://www.ncbi.nlm.nih.gov/geo/query/acc.cgi?acc=gse29385}{GSE29385} & Naim & \shortstack{ Febrile patients with viral \\ infection, mostly influenza } & 80 & 25.0 (IQR 22.0-40.0) & 48 (60) & Illumina & Unk. & 0 (0) & 80 (100) & 0 (0) \\ \hline
      \href{https://www.ncbi.nlm.nih.gov/geo/query/acc.cgi?acc=GSE30119}{GSE30119} & Banchereau & \shortstack{ Pediatric patients with community-acquired \\ bacterial infection } & 81 & 6.5 (IQR 1.92-11.0) & 34 (42) & Illumina & United States of America & 59 (73) & 0 (0) & 22 (27) \\ \hline
      \href{https://www.ncbi.nlm.nih.gov/geo/query/acc.cgi?acc=gse32707}{GSE32707} & Dolinay & \shortstack{ Patients undergoing mechanical ventilation \\ with sepsis or SIRS or neither } & 44 & 56.0 (IQR 45.0-59.0) & 13 (30) & Illumina & United States of America & 0 (0) & 0 (0) & 44 (100) \\ \hline
      \href{https://www.ncbi.nlm.nih.gov/geo/query/acc.cgi?acc=GSE40012}{GSE40012} & Parnell & \shortstack{ Bacterial or influenza A \\ pneumonia or SIRS } & 54 & 59.0 (IQR 46.5-67.0) & 20 (37) & Illumina & Australia & 16 (30) & 8 (15) & 30 (56) \\ \hline
      \href{https://www.ncbi.nlm.nih.gov/geo/query/acc.cgi?acc=GSE40165}{GSE40165} & Nguyen & Febrile children with dengue & 123 & 12.0 (IQR 10.0-14.0) & 85 (69) & Illumina & Vietnam & 0 (0) & 123 (100) & 0 (0) \\ \hline
      \href{https://www.ncbi.nlm.nih.gov/geo/query/acc.cgi?acc=gse40396}{GSE40396} & Hu & \shortstack{ Febrile children with bacterial \\ or viral infection } & 52 & 0.92 (IQR 0.33-1.6) & 17 (33) & Illumina & United States of America & 8 (15) & 22 (42) & 22 (42) \\ \hline
      \href{https://www.ncbi.nlm.nih.gov/geo/query/acc.cgi?acc=GSE40586}{GSE40586} & Lill & Community-acquired bacterial meningitis & 21 & 57.0 (IQR 53.0-70.5) & 0 (0) & Affymetrix & Estonia & 21 (100) & 0 (0) & 0 (0) \\ \hline
      \href{}{INF-02} & Liesenfeld & \shortstack{ Patients with suspected infection/sepsis \\ in ED } & 59 & 39.0 (IQR 29.0-56.0) & 17 (28) & NanoString$^{\text{\textregistered}}$ & USA/Greece & 24 (41) & 7 (12) & 28 (47) \\ \hline
      \href{https://doi.org/10.1038/s41467-020-14975-w}{INF-IIS-01} & Rogers & \shortstack{ ICU patients with at \\ least one risk factor for acute respiratory distress syndrome } & 13 & 57.0 (IQR 42.0-72.0) & 10 (77) & NanoString & USA & 6 (46) & 1 (8) & 6 (46) \\ \hline
      \href{}{INF-IIS-03} & Liesenfeld & \shortstack{ Patients with bacterial upper \\ respiratory tract infections } & 13 & 65.0 (IQR 37.0-85.0) & 8 (62) & NanoString & Greece & 5 (38) & 8 (62) & 0 (0) \\ \hline
      \href{https://doi.org/10.1186/1471-2334-14-272}{INF-IIS-04} & Giamarellos-Bourboulis & \shortstack{ ED and hospitalized sepsis \\ patients } & 3 & 72.0 (IQR 70.5-74.0) & 2 (67) & NanoString & Greece & 3 (100) & 0 (0) & 0 (0) \\ \hline
      \href{https://pubmed.ncbi.nlm.nih.gov/33374697}{INF-IIS-10} & Iglesias & \shortstack{ Sepsis or septic shock \\ patients admitted to ICU } & 6 & 64.5 (IQR 60.25-71.0) & 3 (50) & NanoString & USA & 6 (100) & 0 (0) & 0 (0) \\ \hline
      \href{https://doi.org/10.1097/CCM.0000000000005119}{INF-IIS-11} & Bauer & \shortstack{ Patients with suspected infection/sepsis \\ in ED } & 25 & 77.0 (IQR 65.0-86.0) & 14 (56) & NanoString & Germany & 16 (64) & 5 (20) & 4 (16) \\ \hline
      \href{https://doi.org/10.1111/eci.13626}{INF-IIS-21} & Almansa & \shortstack{ Patients with suspected severe \\ respiratory infection } & 5 & 81.0 (IQR 77.0-82.0) & 5 (100) & NanoString & Spain & 0 (0) & 5 (100) & 0 (0) \\ \hline
      
  \end{tabular}}
  \label{tab:bvn_study_summary}
\end{table}

\begin{table}
  \centering

  \caption{{\bf Development datasets for prognostic classifier.} SJIA:
    Systemic Juvenile Idiopathic Arthritis.}

  \resizebox{\textwidth}{!}{\begin{tabular}{|c|c|c|c|c|c|c|c|c|c|c|}
      \hline
      Study & Author & Description & N & Age$^{*}$ & Male$^{*}$ & Platform & Country & 30-day Mortality & Survival \\ \hline
      \href{http://www.ebi.ac.uk/arrayexpress/experiments/E-MEXP-3589/}{EMEXP3589} & Almansa & \shortstack{ Patients hospitalized with COPD \\ exacerbation } & 27 & Unk. & 16 (59) & Agilent & Spain & 0 (0) & 27 (100) \\ \hline
      \href{http://www.ebi.ac.uk/arrayexpress/experiments/E-MTAB-1548/}{EMTAB1548} & Almansa & \shortstack{ Surgical patients with sepsis \\ (EXPRESS) } & 140 & 72.0 (IQR 61.0-78.0) & 95 (68) & Agilent & Spain & 12 (9) & 128 (91) \\ \hline
      \href{https://www.ebi.ac.uk/biostudies/arrayexpress/studies/E-MTAB-3162}{EMTAB3162} & van den Ham & Patients with dengue & 21 & 20.0 (IQR 17.0-28.0) & 11 (52) & Affymetrix & Indonesia & 0 (0) & 21 (100) \\ \hline
      \href{https://www.ncbi.nlm.nih.gov/geo/query/acc.cgi?acc=GSE42026}{GSE42026} & Herberg & \shortstack{ Severe H1N1/09 influenza or \\ bacterial infection } & 59 & 1.25 (IQR 0.38-4.0) & 33 (56) & Illumina & United Kingdom & 0 (0) & 59 (100) \\ \hline
      \href{https://www.ncbi.nlm.nih.gov/geo/query/acc.cgi?acc=GSE47655}{GSE47655} & Stone & Severe anaphylaxis in ED & 6 & Unk. & 0 (0) & Affymetrix & Australia & 0 (0) & 6 (100) \\ \hline
      \href{https://www.ncbi.nlm.nih.gov/geo/query/acc.cgi?acc=GSE51808}{GSE51808} & Kwissa & Patients with dengue & 28 & Unk. & 0 (0) & Affymetrix & Thailand & 0 (0) & 28 (100) \\ \hline
      \href{https://www.ncbi.nlm.nih.gov/geo/query/acc.cgi?acc=GSE54514}{GSE54514} & Parnell & Sepsis patients in ICU & 19 & Unk. & 0 (0) & Illumina & Unk. & 4 (21) & 15 (79) \\ \hline
      \href{https://www.ncbi.nlm.nih.gov/geo/query/acc.cgi?acc=gse57183}{GSE57183} & Shenoi & Febrile children with SJIA & 11 & 3.6 (IQR 3.3-7.3) & 5 (45) & Illumina & United States of America & 0 (0) & 11 (100) \\ \hline
      \href{https://www.ncbi.nlm.nih.gov/geo/query/acc.cgi?acc=GSE25504}{GSE25504} & Smith & Neonatal sepsis & 33 & 0.0 (IQR 0.0-0.0) & 19 (57) & Affymetrix & United Kingdom & 2 (6) & 31 (94) \\ \hline
      \href{https://www.ncbi.nlm.nih.gov/geo/query/acc.cgi?acc=GSE60244}{GSE60244} & Suarez & Lower respiratory tract infections & 118 & 63.0 (IQR 50.0-77.0) & 37 (31) & Illumina & United States of America & 0 (0) & 118 (100) \\ \hline
      \href{https://www.ncbi.nlm.nih.gov/geo/query/acc.cgi?acc=gse63881}{GSE63881} & Hoang & \shortstack{ Pediatric patients with Kawasaki \\ disease } & 171 & 2.75 (IQR 1.42-4.25) & 102 (60) & Illumina & United States of America & 0 (0) & 171 (100) \\ \hline
      \href{https://www.ncbi.nlm.nih.gov/geo/query/acc.cgi?acc=GSE64456}{GSE64456} & Mahajan & \shortstack{ Febrile infants with bacterial \\ or viral infection in ED } & 200 & 0.1 (IQR 0.06-0.13) & 106 (53) & Illumina & United States of America & 0 (0) & 200 (100) \\ \hline
      \href{https://www.ncbi.nlm.nih.gov/geo/query/acc.cgi?acc=GSE65682}{GSE65682} & Scicluna & \shortstack{ Suspected but negative for \\ CAP } & 106 & Unk. & 0 (0) & Affymetrix & Unk. & 23 (22) & 83 (78) \\ \hline
      \href{https://www.ncbi.nlm.nih.gov/geo/query/acc.cgi?acc=GSE66099}{GSE66099} & Sweeney & \shortstack{ Pediatric ICU patients with \\ sepsis, SIRS } & 229 & 2.45 (IQR 1.0-5.88) & 94 (41) & Affymetrix & United States of America & 28 (12) & 201 (88) \\ \hline
      \href{https://www.ncbi.nlm.nih.gov/geo/query/acc.cgi?acc=GSE67059}{GSE67059} & Heinonen & Children with rhinovirus & 80 & 0.83 (IQR 0.3-1.29) & 53 (66) & Illumina & United States of America & 0 (0) & 80 (100) \\ \hline
      \href{https://www.ncbi.nlm.nih.gov/geo/query/acc.cgi?acc=GSE68310}{GSE68310} & Zhai & \shortstack{ Outpatients with acute respiratory \\ viral infections } & 104 & 20.96 (IQR 20.09-22.76) & 50 (48) & Illumina & United States of America & 0 (0) & 104 (100) \\ \hline
      \href{https://www.ncbi.nlm.nih.gov/geo/query/acc.cgi?acc=GSE72810}{GSE72810} & Herberg & \shortstack{ Pediatric patients with bacterial \\ or viral infection } & 72 & 1.83 (IQR 0.88-3.29) & 8 (11) & Illumina & United Kingdom & 0 (0) & 72 (100) \\ \hline
      \href{https://www.ncbi.nlm.nih.gov/geo/query/acc.cgi?acc=GSE73461}{GSE73461} & Wright & \shortstack{ Pediatric patients with bacterial \\ or viral infection or other inflammatory diseases } & 404 & 2.79 (IQR 0.92-8.81) & 165 (41) & Illumina & United Kingdom & 0 (0) & 404 (100) \\ \hline
      \href{https://www.ncbi.nlm.nih.gov/geo/query/acc.cgi?acc=gse77087}{GSE77087} & de Steenhuijsen Piters & \shortstack{ Infants with respiratory synctial \\ virus } & 41 & 0.45 (IQR 0.14-0.69) & 25 (61) & Illumina & United States of America & 0 (0) & 41 (100) \\ \hline
      \href{https://www.ncbi.nlm.nih.gov/geo/query/acc.cgi?acc=GSE82050}{GSE82050} & Tang & \shortstack{ Moderate and severe influenza \\ infection } & 24 & 64.5 (IQR 48.5-74.25) & 13 (54) & Agilent & Germany & 0 (0) & 24 (100) \\ \hline
      \href{https://www.ncbi.nlm.nih.gov/geo/query/acc.cgi?acc=GSE95233}{GSE95233} & Venet & Septic shock & 51 & 66.0 (IQR 53.5-73.5) & 31 (61) & Affymetrix & France & 17 (33) & 34 (67) \\ \hline
      \href{https://www.ncbi.nlm.nih.gov/geo/query/acc.cgi?acc=gse103842}{GSE103842} & Rodriguez-Fernandez & \shortstack{ Infants with respiratory synctial \\ virus } & 62 & 0.25 (IQR 0.17-0.44) & 39 (63) & Illumina & United States of America & 0 (0) & 62 (100) \\ \hline
      \href{https://www.ncbi.nlm.nih.gov/geo/query/acc.cgi?acc=GSE13015}{GSE13015gpl6102} & Pankla & \shortstack{ Sepsis, many cases from \\ burkholderia } & 55 & 54.0 (IQR 48.0-61.0) & 26 (47) & Illumina & Thailand & 15 (27) & 40 (73) \\ \hline
      \href{https://www.ncbi.nlm.nih.gov/geo/query/acc.cgi?acc=GSE13015}{GSE13015gpl6947} & Pankla & \shortstack{ Sepsis, many cases from \\ burkholderia } & 15 & 49.0 (IQR 43.5-59.5) & 6 (40) & Illumina & Thailand & 7 (47) & 8 (53) \\ \hline
      \href{https://www.ncbi.nlm.nih.gov/geo/query/acc.cgi?acc=GSE22098}{GSE22098} & Berry & \shortstack{ Bacterial infection or other \\ inflammatory conditions } & 193 & 16.0 (IQR 11.0-26.0) & 59 (31) & Illumina & Unk. & 0 (0) & 193 (100) \\ \hline
      \href{https://www.ncbi.nlm.nih.gov/geo/query/acc.cgi?acc=GSE27131}{GSE27131} & Berdal & Severe H1N1 & 7 & 38.0 (IQR 33.0-50.0) & 6 (86) & Affymetrix & Norway & 2 (28) & 5 (71) \\ \hline
      \href{https://www.ncbi.nlm.nih.gov/geo/query/acc.cgi?acc=gse28991}{GSE28991} & Naim & Patients with suspected dengue & 11 & Unk. & 0 (0) & Illumina & Unk. & 0 (0) & 11 (100) \\ \hline
      \href{https://www.ncbi.nlm.nih.gov/geo/query/acc.cgi?acc=GSE30119}{GSE30119} & Banchereau & \shortstack{ Pediatric patients with community-acquired \\ bacterial infection } & 59 & 6.5 (IQR 1.92-11.0) & 34 (57) & Illumina & United States of America & 0 (0) & 59 (100) \\ \hline
      \href{https://www.ncbi.nlm.nih.gov/geo/query/acc.cgi?acc=gse32707}{GSE32707} & Dolinay & \shortstack{ Patients undergoing mechanical ventilation \\ with sepsis or SIRS or neither } & 69 & 56.0 (IQR 45.0-59.0) & 13 (19) & Illumina & United States of America & 25 (36) & 44 (64) \\ \hline
      \href{https://www.ncbi.nlm.nih.gov/geo/query/acc.cgi?acc=GSE40012}{GSE40012} & Parnell & \shortstack{ Bacterial or influenza A \\ pneumonia or SIRS } & 39 & 59.0 (IQR 46.5-67.0) & 20 (51) & Illumina & Australia & 5 (13) & 34 (87) \\ \hline
      \href{https://www.ncbi.nlm.nih.gov/geo/query/acc.cgi?acc=GSE40165}{GSE40165} & Nguyen & Febrile children with dengue & 123 & 12.0 (IQR 10.0-14.0) & 85 (69) & Illumina & Vietnam & 0 (0) & 123 (100) \\ \hline
      \href{https://www.ncbi.nlm.nih.gov/geo/query/acc.cgi?acc=gse40396}{GSE40396} & Hu & \shortstack{ Febrile children with bacterial \\ or viral infection } & 30 & 0.92 (IQR 0.33-1.6) & 17 (56) & Illumina & United States of America & 0 (0) & 30 (100) \\ \hline
      \href{https://www.ncbi.nlm.nih.gov/geo/query/acc.cgi?acc=GSE40586}{GSE40586} & Lill & Community-acquired bacterial meningitis & 21 & 57.0 (IQR 53.0-70.5) & 0 (0) & Affymetrix & Estonia & 2 (10) & 19 (90) \\ \hline
      \href{}{INF-02} & Liesenfeld & \shortstack{ Patients with suspected infection/sepsis \\ in ED } & 61 & 41.0 (IQR 29.0-57.0) & 18 (30) & NanoString$^{\text{\textregistered}}$ & USA/Greece & 2 (3) & 59 (97) \\ \hline
      \href{https://doi.org/10.1038/s41467-020-14975-w}{INF-IIS-01} & Rogers & \shortstack{ ICU patients with at \\ least one risk factor for acute respiratory distress syndrome } & 11 & 57.0 (IQR 39.5-81.5) & 10 (91) & NanoString & USA & 5 (45) & 6 (55) \\ \hline
      \href{}{INF-IIS-03} & Liesenfeld & \shortstack{ Patients with bacterial upper \\ respiratory tract infections } & 19 & 77.0 (IQR 51.5-85.0) & 12 (63) & NanoString & Greece & 9 (47) & 10 (53) \\ \hline
      \href{https://doi.org/10.1186/1471-2334-14-272}{INF-IIS-04} & Giamarellos-Bourboulis & \shortstack{ ED and hospitalized sepsis \\ patients } & 3 & 72.0 (IQR 70.5-74.0) & 2 (67) & NanoString & Greece & 1 (33) & 2 (67) \\ \hline
      \href{https://pubmed.ncbi.nlm.nih.gov/33374697}{INF-IIS-10} & Iglesias & \shortstack{ Sepsis or septic shock \\ patients admitted to ICU } & 6 & 64.5 (IQR 60.25-71.0) & 3 (50) & NanoString & USA & 1 (17) & 5 (83) \\ \hline
      \href{https://doi.org/10.1097/CCM.0000000000005119}{INF-IIS-11} & Bauer & \shortstack{ Patients with suspected infection/sepsis \\ in ED } & 25 & 77.0 (IQR 65.0-86.0) & 14 (56) & NanoString & Germany & 4 (16) & 21 (84) \\ \hline
      \href{https://doi.org/10.1111/eci.13626}{INF-IIS-21} & Almansa & \shortstack{ Patients with suspected severe \\ respiratory infection } & 5 & 81.0 (IQR 77.0-82.0) & 5 (100) & NanoString & Spain & 0 (0) & 5 (100) \\ \hline
      
  \end{tabular}}
  \label{tab:sev_study_summary}
\end{table}

\subsection{Feature selection}
\label{FEATURE_SELECTION}

\begin{figure}
  \centering
  \includegraphics[width=1\textwidth]{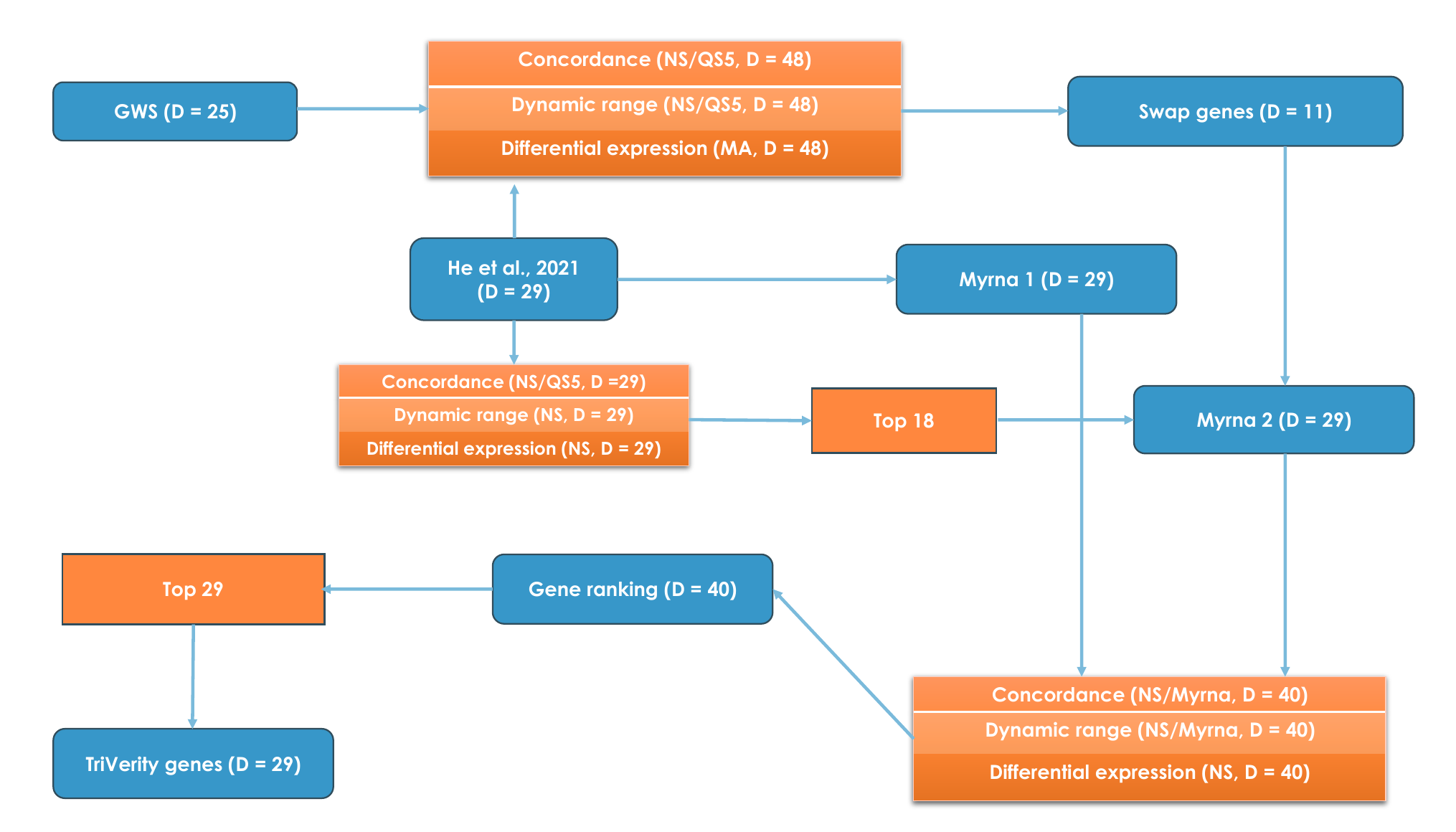}

  \caption{{\bf Overview of selection of gene features compatible with
      LAMP amplification.} GWS: Genome-wide search; MA: microarray;
    NS: NanoString$^{\text{\textregistered}}$.}

  \label{fig:feature_selection}
\end{figure}

Feature selection is illustrated in
Fig.~\ref{fig:feature_selection}. It was assay-guided in the sense
that all selected features must have exhibited acceptable
amplification by LAMP.

\begin{enumerate}
  
\item The ``benchtop LAMP'' experiments revealed lack of amplification
  of 10 genes, which were replaced by 10 genes with a good trade-off
  between the LAMP amplification and clinical
  performance~[\citenum{HE2021}].

\item Subsequent experiments on the target platform revealed lack of
  amplification of additional 11 genes, due to differences between the
  benchtop and Myrna\textsuperscript{\texttrademark}
  amplification. The non-amplified genes were replaced with genes
  which had good trade-off between Myrna amplification and clinical
  performance using the process shown in
  Fig.~\ref{fig:feature_selection}. Briefly, we started with a new
  sorted list of predictive and prognostic genes identified using
  midplex-genome-wide search for biological signal. We then went down
  the list and synthesized Myrna primers (chemicals used to detect
  mRNA), where possible.  The highest ranked gene with amplified
  primer replaced the highest ranked gene without the amplification in
  the existing signature. The ranking was based on a heuristic
  function comprising Shapley value for the gene and concordance with
  the NanoString platform. The procedure was repeated until all
  non-amplified genes were replaced.

\item Following the transfer of the gene set to Myrna, we attempted
  reduction of the signature size using a greedy forward search.

\end{enumerate}

\begin{figure}[ht]
\centering

\begin{subfigure}[b]{1\textwidth}
  \centering
  \includegraphics[width=1\textwidth]{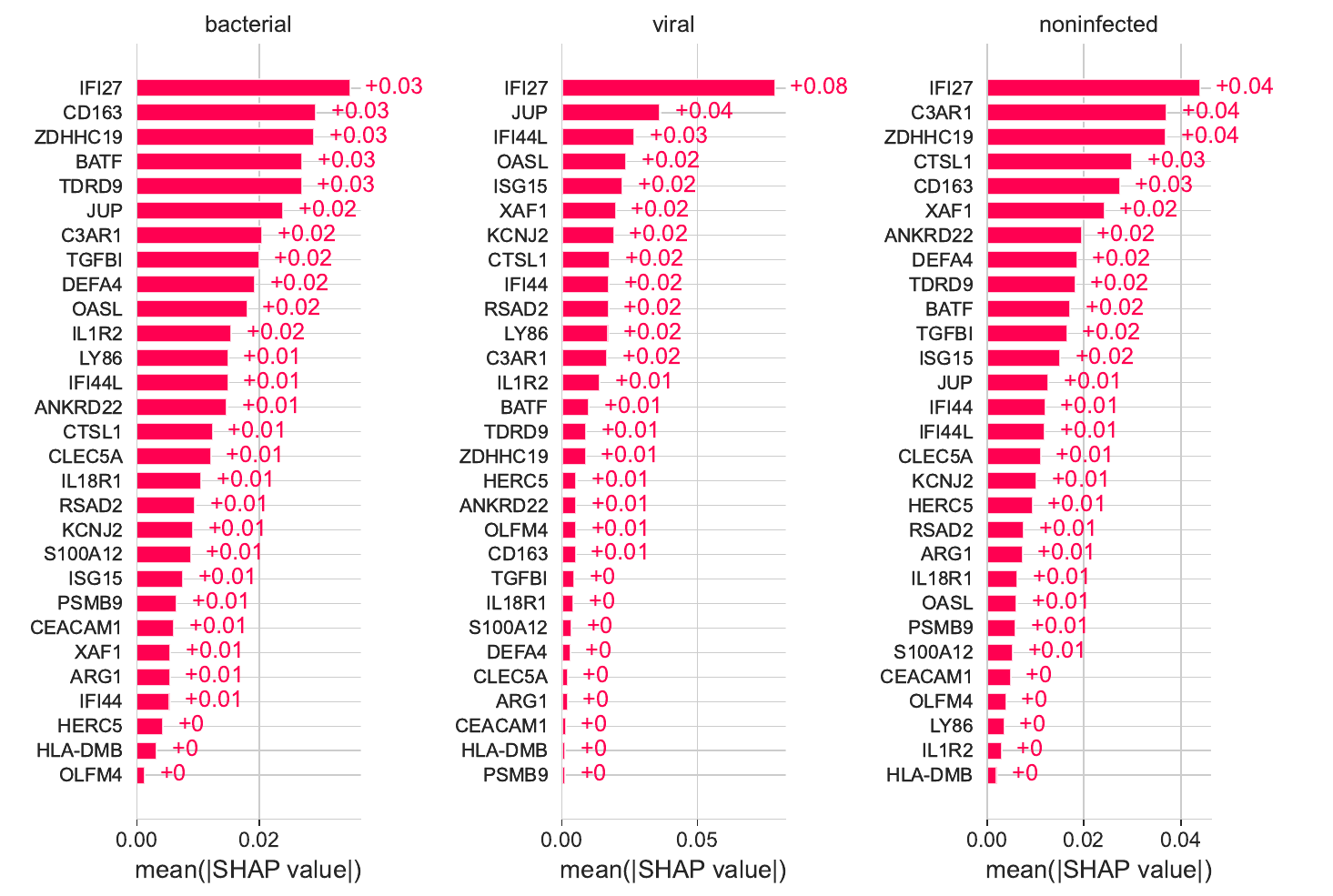}
  \caption{}
\end{subfigure}
\vspace{1cm} 
\begin{subfigure}[b]{0.8\textwidth}
  \centering
  \includegraphics[width=0.8\textwidth]{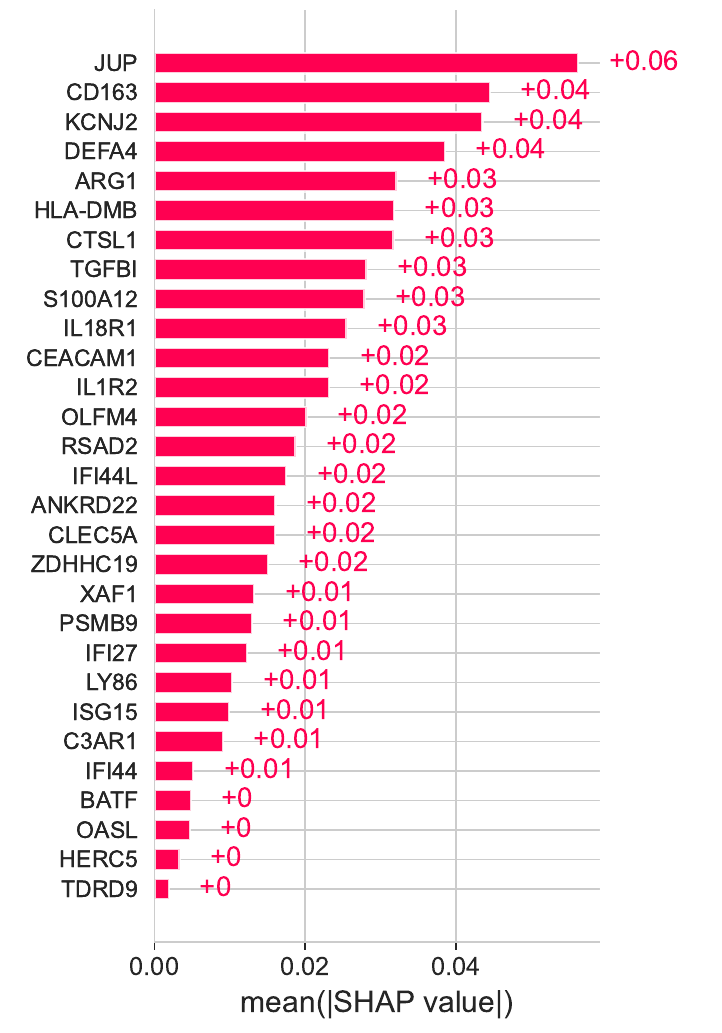}
  \caption{}
\end{subfigure}

\caption{{\bf TriVerity\textsuperscript{\texttrademark} genes, ordered
    by absolute values of Shapley values for each probability.} {\bf
    a,} BVN classifier. {\bf b,} SEV classifier.}

\label{fig:genes}

\end{figure}

The final set of genes chosen by the feature selection process is
shown in Fig.~\ref{fig:genes}.

\subsection{Concordance Filtering ablation}
\label{CF_ABLATION}

To estimate the impact of CF, we performed an ablation study by
evaluating performance of the leading classifiers (MLP and LOGR) with
and without the CF. Table~\ref{tab:cf_ablation} shows that CF
significantly improves the PCC for MLP classifier, without loss of
accuracy.

\begin{table}

  \caption{{\bf Impact of CF on BVN classifier accuracy and
      concordance.} The CF experiments used the filtering value of
    0.92 Pearson correlation coefficient between probabilities
    assigned to replicate samples.}

  \label{tab:cf_ablation}
  \centering
  \begin{tabular}{cccc}
    \toprule
    Algorithm    & Statistic           & With CF                & Without CF            \\
    \midrule
    LOGR         & CV mAUROC           & 0.886                  & 0.885                 \\
    LOGR         & Validation mAUROC   & 0.834                  & 0.833                 \\
    LOGR         & Bacterial PCC       & 0.938                  & 0.939                 \\
    LOGR         & Viral PCC           & 0.962                  & 0.963                 \\
    MLP          & CV mAUROC           & 0.895                  & 0.897                 \\
    MLP          & Validation mAUROC   & 0.832                  & 0.834                 \\
    MLP          & Bacterial PCC       & 0.942                  & 0.891                 \\
    MLP          & Viral PCC           & 0.960                  & 0.940                 \\
    \bottomrule
  \end{tabular}
\end{table}

\end{document}